\documentclass[a4paper,10pt]{article}
\usepackage{graphicx}
\usepackage{amsmath}
\title{First-passage problems in DNA replication: effects of template tension on stepping and exonuclease activities of a DNA polymerase motor}
\author{Ajeet K. Sharma and Debashish Chowdhury\\
Department of Physics, Indian Institute of Technology,\\
 Kanpur, 208016}
\begin{document}
\maketitle

\begin{abstract}
A DNA polymerase (DNAP) replicates a template DNA strand. It also
exploits the template as the track for its own motor-like mechanical 
movement. In the polymerase mode it elongates the nascent DNA by one
nucleotide in each step. But, whenever it commits an error by 
misincorporting an incorrect nucleotide, it can switch to an 
exonuclease mode. In the latter mode it excises the wrong nucleotide 
before switching back to its polymerase mode. We develop a stochastic 
kinetic model of DNA replication that mimics an {\it in-vitro} 
experiment where a single-stranded DNA, subjected to a mechanical 
tension $F$, is converted to a double-stranded DNA by a single DNAP. 
The $F$-dependence of the average rate of replication, which depends 
on the rates of both polymerase and exonuclease activities of the 
DNAP, is in good 
qualitative agreement with the corresponding experimental results. 
We introduce 9 novel distinct {\it conditional dwell times} of a DNAP. 
Using the methods of first-passage times, we also derive the exact 
analytical expressions for the probability distributions of these 
conditional dwell times. The predicted $F$-dependence of these 
distributions are, in principle, accessible to single-molecule experiments.
\end{abstract}

\section{Introduction}
\label{sec-introduction}

A linear molecular motor is either a macromolecule or macromolecular complex 
that moves along a filamentous track \cite{howard01a,bustamante01,fisher07,chowdhury13a,chowdhury13b}. 
In spite of its noisy stepping kinetics, on the average, it moves in a 
directed manner. Its mechanical work is fuelled by the input energy which, 
for many motors, is chemical energy. The distributions of the dwell times of 
a motor at discrete positions on its track as well as the duration of many 
complex motor-driven intracellular processes have been calculated 
\cite{kolomeisky05,liao07,linden07,tsygankov07,chemla08,moffitt10b,sharma11a} 
using the methods of first-passage times \cite{redner}, a well-known formalism 
in non-equilibrium statistical mechanics. 
Experimentally measured distributions of dwell times of a motor can be utilized to extract useful information on its kinetic scheme \cite{moffitt10b,schnitzer95}.

For motors which can step both forward and backward on a linear track, four 
distinct conditional dwell times can be defined; distributions of these four 
conditional dwell times have been calculated for some motors 
\cite{linden07,tsygankov07,garai11,sharma11a}. 
In this paper we consider a specific molecular motor called DNA polymerase 
(DNAP) and argue that its movements on the track is characterized by 
{\it nine} distinct conditional dwell times because of the coupling of its 
dual roles during its key biological function. We define these nine conditional 
dwell times and calculate their distributions analytically treating each of 
these as an appropriate first-passage time. As a byproduct of this exercise, 
we obtain an important result, namely, its mean velocity, that characterizes 
one of its average properties; the theoretically predicted behaviour is 
consistent with the corresponding experimental observations reported earlier 
in the literature. The distributions of the 
nine conditional dwell times are new predictions which, we believe, can be 
tested by single-molecule experiments.

Deoxyribonucleic acid (DNA) is a polynucleotide, i.e., a linear heteropolymer
whose monomeric subunits are drawn from a pool of four different species of 
nucleotides, namely, A (Adenine), T (Thymine), C (Cytosine) and G (Guanine). 
In this heteropolymer the nucleotides are linked by phosphodiester bonds. 
The genetic message is chemically encoded in the sequence of the nucleotide 
species. DNA polymerase (DNAP)
\cite{kornberg92}, 
the enzyme that replicates DNA, carries out a template-directed polymerization 
\cite{sharma12}.
During this processes, repetitive cycles of nucleotide selection and 
phosphodiester bond formation is performed to polymerize a DNA strand.
In every elongation cycle, hydrolysis of the substrate molecule supplies
sufficient amount of energy to the DNAP for performing its function.
Therefore, DNAPs are also regarded as molecular motor 
\cite{howard01a,bustamante01,fisher07,chowdhury13a,chowdhury13b}
that transduce chemical energy into mechanical work while translocating 
step-by-step on the template DNA strand that serves as a track for these 
motors.

In an {\it in-vitro} experiment, Wuite et al. \cite{wuite00} applied a
tension on a ssDNA. The two ends of this DNA fragment were connected to
two dielectric beads; one end was held by micro-pipette, while the other 
end, trapped optically by a laser beam, was pulled. This DNA fragment 
also served as a template for the replication process carried out by 
a DNAP. Replication converted the ssDNA into a dsDNA. The average rate of 
replication was found to vary {\it nonmonotonically} with the tension applied 
on the template strand \cite{wuite00}. Similar results were obtained also 
in the experiments carried out by Maier et al. \cite{maier00}, where 
magnetic tweezers were used to apply the tension on template DNA. 
The observed nonmonotonic variation of the average rate of replication
was explained \cite{wuite00,maier00,goel01,goel03,andricioaei04} 
as a consequence of the difference in the force-extension 
curves of ssDNA and dsDNA \cite{rouzina01}. 

Fidelity of replication carried out by a DNAP is normally very high 
\cite{kunkel09}. It achieves such high accuracy by discriminating 
between the correct and incorrect nucleotides by {\it kinetic 
proofreading}. The mechanism of kinetic proofreading enables the DNAP 
to reduce the error ratio to values far lower than the thermodynamically 
allowed value of $exp(-\dfrac{\Delta {\cal F}}{k_{B}T})$, where 
$\Delta {\cal F}$ is the free energy difference  of enzyme substrate 
complex for correct and incorrect nucleotides. Thus, DNAP is capable of 
correcting most of its own error during the ongoing replication process itself.

A DNAP performs its normal function as a polymerase by catalyzing the 
elongation of a new ssDNA molecule using another ssDNA as a template. 
However, upon committing a misincorporation of a nucleotide in the 
elongating DNA, the DNAP can detect its own error and transfer the nascent 
DNA to another site where it catalyses excision of the wrongly incorporated 
nucleotide. The distinct sites, where 
the polymerisation (pol) and exonuclease (exo) reactions are catalyzed, 
are separated by 3-4 nm on the same DNAP \cite{ibarra09}. The nascent DNA is transferred 
to the pol site from the exo site after the wrong nucleotide is cleaved from its 
tip by the DNAP. Thus, the transfer of the DNA between the pol and exo sites 
couples the polymerase and exonuclease activities of the DNAP. 

In the next section we develop a microscopic model for the replication of a 
ssDNA template that is subjected to externally applied tension $F$, a 
situation that is very similar to the {\it in-vitro} experiment reported in 
refs. \cite{wuite00,maier00}. 
The rates of both pol and exo activities of the DNAP enter into the 
expression that we derive for the average rate of elongation of the DNA.
The $F$-dependence of this rate is consistent with the experimental 
observations reported in \cite{wuite00,maier00}. 
We then define 9 distinct conditional dwell times of the DNAP and identifying 
each of these with an appropriate first-passage time \cite{redner}, we 
calculate their distributions analytically. 
We believe that experimental measurements of these distributions are likely 
to elucidate the nature of the interplay of the pol and exo activities of 
DNAP.

\section{Model}

\begin{figure}[ht]
\begin{center}
\includegraphics[angle=0, width=0.8 \columnwidth]{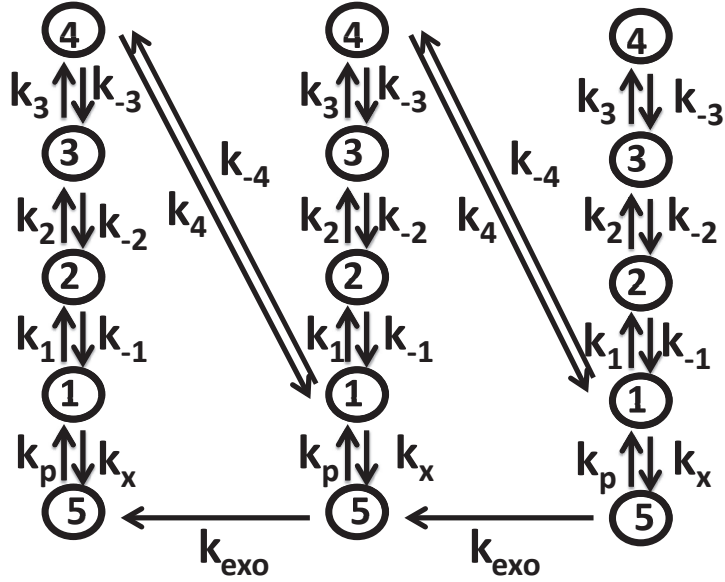}
\end{center}
\caption{A pictorial depiction of 5 state kinetic model for DNA polymerase (see the text for a detailed explanation).}
\label{fig1}
\end{figure}

The nucleotides on the template DNA are labelled sequentially by the integer index $j$ ($j=1,2,...,L$) which also serves to indicate the position of the DNAP on its track. The chemical (or conformtional) state of the DNAP is denoted by a discrete variable $\mu$ ($\mu=1,2...,5$).  The state of the DNAP is during replication is described by the pair $j,\mu$. The kinetic scheme used for our model is adapted from that proposed originally by Patel et al. \cite{patel92} and subsequently utilized by various other groups \cite{xie07,goel03}. The kinetic scheme of our model is shown in figure (\ref{fig1}), where the four different values $1$, $2$, $3$ and $4$ of $\mu$ are the allowed chemical states in the polymerase-active mode of the enzyme, while in chemical state $5$ the exonuclease catalytic site is activated. 

The structure of DNA polymerase resembles a ``cupped right hand'' of a
human, where its sub domains are recognized as palm, thumb and finger 
sub domains \cite{cramer02}. Template DNA enters from the finger sub-domain
and takes exit from thumb sub-domain. The catalytic site where the binding
occurs is located between finger and palm domain. 
Transitions between polymerase activated kinetic states
of the enzyme (i.e., chemical states 1,2, 3 and 4) can be summarized 
as \cite{johnson10,berdis09}
\begin{equation}
E_{o}D_{j}+dNTP \mathop{\rightleftharpoons}^{k_{1}}_{k_{-1}} E_{o}D_{j}dNTP \mathop{\rightleftharpoons}^{k_{2}}_{k_{-2}} E_{c}D_{j}dNTP \mathop{\rightleftharpoons}^{k_{3}}_{k_{-3}}
 E_{c}D_{j+1}PP_{i} \mathop{\rightleftharpoons}^{k_{4}}_{k_{-4}} E_{o}D_{j+1}
\end{equation}
where $E_{c}$ and $E_{o}$ represent the closed and open finger configuration 
DNAP, respectively, while $D_{j}$ denotes the length of the nascent  DNA strand.

Let us start with the state $E_{o}D_{j}$, labelled by $\mu=1$, in which the finger domain of DNAP is open and the DNAP is located at the site $j$ on its template. Now a substrate
molecule (dNTP) binds with the DNAP and resulting state $E_{o}D_{j}dNTP$ is labeled by ``2'' 
. The transition $1 \to 2$ take place with rate $k_{1}$, while corresponding
reverse transition $2 \to 1$ occurs with rate $k_{-1}$. Binding energy of dNTP switches the
open finger configuration of DNAP into closed finger configuration and the 
corresponding transition 2 $(E_{o}D_{j}dNTP)$ $\rightarrow$ 3 $(E_{c}D_{j}dNTP)$
take place at the rate $k_{2}$. The reverse transition $3 \to 2$ occurs at the rate $k_{-2}$.
This new closed finger configuration of DNAP catalyzes the formation of
phosphodiester bond between dNTP and nascent DNA strand thereby elongating 
the nascent DNA from length $j$ to $j+1$; this process is represented by 
the transition  3 $(E_{c}D_{j}dNTP)$ $\to$ 4 ($E_{c}D_{j+1}PP_{i}$) that occurs at the rate $k_{3}$ ($k_{-3}$ being the rate of the reverse transition). Finally, the transition $4(j)$ $\rightarrow$ $1(j+1)$ completes one elongation cycle; the corresponding rates of the forward and reverse transitions are $k_{4}$ and $k_{-4}$, respectively. The 
transition $4(j)$ $\rightarrow$ $1(j+1)$ captures more than one sub-step which includes opening of the finger domain, release of $PP_{i}$ and the forward 
movement of the DNAP to the next site on the template.
   
Immediately after completing one elongation cycle, the DNAP is normally ready to bind with a new substrate molecule and initiate the next elongation cycle. However, if a wrong nucleotide is incorporated in an elongation cycle, the DNAP is likely to transfer the nascent DNA from the pol site to the exo site. This switching from pol to exo activity is represented by the transition $1 \to 5$ which occurs at the rate $k_x$; the reverse transition, without cleavage, takes place at the rate $k_p$. In the exo mode the cleavage of the last incorporated nucleotide, at the rate $k_{exo}$, effectively alters the position of the DNAP from $j+1$ to $j$.

\subsection{Force dependent chemical steps}
\label{sec-forcevel}
External load force tilts the free energies and alters the barriers 
for the forward and reverse transitions \cite{bustamante04}. But,
not all the rate constants change significantly with the tension $F$ 
applied on the template. 
We hypothesize that only the following transitions are affected by the 
tension $F$: 
(I) 3 $\rightarrow$ 4, i.e., the polymerization step, where new dNTP subunit is incorporated into nascent DNA chain and a single stranded nucleotide is converted into a double stranded DNA. 
(II) 1 $\rightarrow$ 5 i.e., the transfer of the nascent DNA from the pol site to the exo site of the DNAP. These two catalytic sites are separated by 3.5 nm and a transfer of the nascent DNA between them includes major change in the DNAP conformation that involves a $\beta$ hairpin \cite{beese93,wang97,shamoo99}. Moreover, polymerase to exonuclease switching causes local melting of the dsDNA. 

Suppose $\Delta \Phi (F)$ is the {\it change} in the free energy barrier 
so that  
\begin{eqnarray}
k_{3}(F)&=&k_{3}(0)exp(-\theta \Delta \Phi/k_{B}T), ~~ k_{-3}(F) = k_{-3}(0)exp ((1-\theta) \Delta \Phi/k_{B}T) \nonumber \\
k_{p}(F)&=&k_{p}(0)exp(-\theta' \Delta \Phi_{x}/k_{B}T), ~~ k_{x}(F) = k_{x}(0)exp ((1-\theta') \Delta \Phi_{x}/k_{B}T) \nonumber \\
\label{eq-fishkolo} 
\end{eqnarray}
where $k_B$ is the Boltzmann constant, $T$ is the absolute temperature 
and $k_{3}(0)$, $k_{-3}(0)$, $k_{x}(0)$, $k_{p}(0)$ are the values of the 
corresponding rate constants in the absence of external force. 
The symbols $\theta (0 \leq \theta \leq 1)$ and 
$\theta' (0 \leq \theta' \leq 1)$ in eqn.(\ref{eq-fishkolo}) are the 
load-sharing parameters \cite{fisher07}.
Note that detailed balance is satisfied by our choice of the 
force-dependence of the rate constants 
when it is satisfied by the corresponding rates in the absence of the force. 
The expressions for $\Delta \Phi(F)$ and $\Delta \Phi_{x}(F)$ are derived 
in appendix A by relating these to $\Delta \Phi' (F)$ which is the change 
in the stretching free energy when a ssDNA is converted into dsDNA.  
As we show in the next section, following force dependence of $k_{3}(F)$ and 
$k_{x}(F)$, 
\begin{equation}
k_{3}(F)= k_{3}(0) exp(- \Delta \Phi(F)/k_{B}T) ~{\rm and}~  k_{x}(F)=k_{x}(0)exp(\Delta \Phi_{x}(F)/k_{B}T), 
\label{eq-fdeprate}
\end{equation} 
together with $k_{-3}(F)=k_{-3}(0), k_{p}(F)=k_{p}(0)$, 
i.e., $\theta=1, \theta'=0$, shows a good qualitative agreement with the 
experimental data. 

\section{Results} 
\label{sec-results}

\subsection{Force velocity curve} 
\label{sec-fvcurve}

In this subsection we derive the force-velocity curve for our model DNAP motor and compare it with those reported earlier in the literature. Let $P_{\mu}(j,t)$ be the probability of finding DNAP in chemical state $\mu$, at the position $j$ on its track, at time $t$. 
The probability to finding the DNA polymerase in chemical state $\mu$, irrespective of its position, is 
\begin{equation}
P_{\mu}(t)=\sum_{j=1}^{L} P_{\mu}(j,t)
\end{equation} 
where $L$ is the total number of nucleotides in template DNA strand.
Normalisation of the probability imposes the condition
\begin{equation}
\sum_{\mu=1}^{5} P_{\mu}(t)=1
\end{equation} 
at all times.
The time evolution of the probability $P_{\mu}(t)$ is governed by following equations
\begin{equation}
\dfrac{dP_{1}(t)}{dt}=-(k_{x}+k_{1}+k_{-4})P_{1}(t)+k_{-1}P_{2}(t)+k_{4}P_{4}(t)+k_{p}P_{5}(t)
\end{equation}
\begin{equation}
\dfrac{dP_{2}(t)}{dt}=k_{1}P_{1}(t)-(k_{-1}+k_{2})P_{2}(t)+k_{-2}P_{3}(t)
\end{equation}
\begin{equation}
\dfrac{dP_{3}(t)}{dt}=k_{2}P_{2}(t)-(k_{-2}+k_{3})P_{3}(t)+k_{-3}P_{4}(t)
\end{equation}
\begin{equation}
\dfrac{dP_{4}(t)}{dt}=k_{-4}P_{1}(t)+k_{3}P_{3}(t)-(k_{-3}+k_{4})P_{4}(t)
\end{equation}
\begin{equation}
\dfrac{dP_{5}(t)}{dt}=k_{x}P_{4}(t)-k_{p}P_{5}(t)
\end{equation}

Now we solve these equations in steady state and calculate the 
probability of finding the DNA polymerase in $\mu$th chemical state ($P_{\mu}^{st}$).
\begin{equation}
P_{\mu}^{st}=\dfrac{x_{\mu}}{x_{1}+x_{2}+x_{3}+x_{4}+x_{5}}
\end{equation}
Expressions for $x_{\mu}$'s are given in Appendix B.

Now we define the average rate of polymerization $V_{p}$ and the average rate of excision $V_{e}$ as
\begin{equation}
V_{p}= P_{1}^{st}k_{1}-P_{2}^{st}k_{-1} ~{\rm and}~  V_{e} = k_{exo}P_{5}^{st}
\end{equation}
Therefore, the average velocity of the DNAP on its track is
\begin{equation}
V=V_{p}-V_{e}
\end{equation}

\begin{figure}
\begin{center}
\includegraphics[angle=-90,width=0.8\columnwidth]{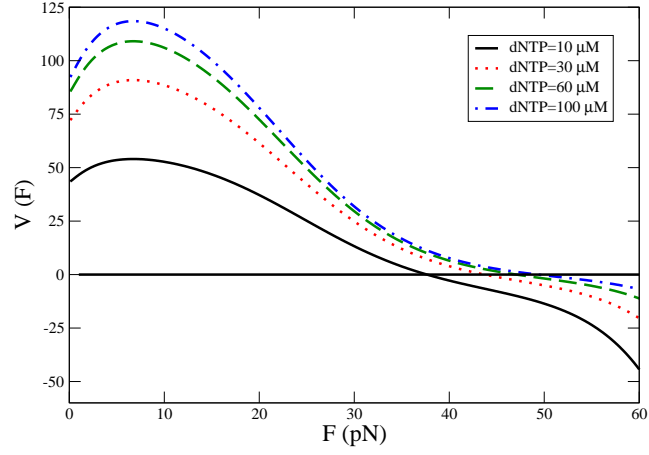}
\end{center}
\caption{Velocity of DNA polymerase is plotted against the force
applied on template strand for a few different values of dNTP 
concentration. The numerical values of the parameters used for 
this plot are listed in table \ref{tab-rateconst}.}
\label{force1}
\end{figure}
In figure (\ref{force1}) the average velocity of the  DNAP is plotted
against the tension applied on DNA track. Rate constants used for this
plot are collected from the literature \cite{patel92} and listed in 
table \ref{tab-rateconst}.\\
\begin{table}
\begin{tabular}{|l|l|}
  \hline
  Rate constant & Numerical value \\ \hline 
  \hline
  $k_{1}$ & 50 $\mu M^{-1}s^{-1}$ \\ \hline
  $k_{-1}$ & 1000 $s^{-1}$ \\  \hline
  $k_{2}$ & 300 $s^{-1}$ \\  \hline
  $k_{-2}$ & 100 $s^{-1}$ \\  \hline
  $k_{3}(0)$ & 9000 $s^{-1}$ \\  \hline
  $k_{-3}$ & 18000 $s^{-1}$ \\  \hline
  $k_{4}$ & 600 $s^{-1}$ \\   \hline
  $k_{-4}$ & 25 $s^{-1}$ \\   \hline
  $k_{x}(0)$ & .2 $s^{-1}$ \\   \hline
  $k_{p}$ & 700 $s^{-1}$ \\   \hline
  $k_{exo}$ & 900 $s^{-1}$ \\
  \hline
  \end{tabular} 
\caption{Numerical values of the rate constants used for 
graphical plotting of some typical curves obtained from 
the analytical expressions derived in this paper.} 
\label{tab-rateconst}
\end{table}
\\  
Because of the $F$-dependence of the form assumed in (\ref{eq-fdeprate}),
at lower tension transition 2 $\rightarrow$ 3 is rate limiting while at 
higer values of  tension  3 $\rightarrow$ 4 becomes the rate limiting step. 
Frequent $poly$ $\rightarrow$ $exo$ switching cause the significant 
increase in the exonuclease cleaving at higher forces. Observed trend of 
variation of the average velocity is the direct consequence of the 
nonmonotonic behavior of the $\Delta \Phi (F)$, shown in figure (\ref{phie}).

\subsection{Distributions of dwell times and exonuclease turnover times}

The average velocity of a DNAP and its dependence on the tension applied on the corresponding template does not provide any information on the intrinsic fluctuations in 
both the pol and exo activities of these machines. Probing fluctuations in the kinetics of molecular machines have become possible  because of the recent advances in single molecule imaging, manipulation and enzymology. 
In this section we investigate theoretically how the fluctuations in the pol and exo activities of a DNAP would vary with the tension applied on the template DNA. For this purpose we use the same kinetic model introduced in section \ref{sec-introduction}, that we have used in subsection \ref{sec-forcevel} for calculating the average properties of DNAP.
  
The variable chosen to characterize the 
fluctuations in replication process is the time of dwell of DNAP at a single
nucleotide on the template, which is nothing but the effective duration of its stay in that location. 
While moving on the one dimensional template strand three different mechanical
steps are taken by DNAP, which are \\ 
(1) Forward step in the pol mode: $4(j)$ $\rightarrow$ $1(j+1)$.\\
(2) Backward step in the pol mode: $1(j+1)$ $\rightarrow$ $4(j)$.\\
(3) Backward step (caused by cleavage) in the exo mode: $5(j+1)$ $\rightarrow$ $5(j)$.

If a molecular motor takes more than one type of mechanical step then the fluctuations in the durations of its dwell at different locations cannot be characterized by a single distribution; instead, distributions of more than one type of conditional dwell times can be defined \cite{chemla08}. So, in the context of our model of DNAP, three different types of mechanical step would generate nine different distribution of conditional dwell times. We denote the  forward, backward and cleavage steps are by the symbols $+$, $-$ and $x$, respectively. $\Psi_{mn} (t)$ is the conditional dwell time of the DNA polymerase
when step m is followed by n, where the three allowed values of each of the subscripts $m$ and $n$ are $+,-,x$.
For the convenience of calculation of the distributions  $\Psi_{mn} (t)$, first we assume that the DNAP is already at the $j_{th}$ site on the template strand and that the rate constants for all the transitions leading to this special site $j$ are equated to zero. In other words, \\
(1) $k_{4}=0$ only for the transition $4(j-1) \rightarrow 1(j)$ (and not for any $i \neq j$),\\
(2) $k_{-4}=0$ only for $1(j+1) \rightarrow 4(j)$ (and not for any $i \neq j$),\\
(3) $k_{exo}=0$ only for $5(j+1) \rightarrow 5(j)$ (and not for any $i \neq j$).\\
Now appropriate initial conditions will ensure the type of previous step taken by DNAP.\\ 

If $P_{\mu}(j,t)$ is the probability of finding the DNA polymerase in $\mu_{th}$ chemical
state at site $j$ at time $t$, then time evolution of these probabilities are governed
by following master equation.
\begin{equation}
\dfrac{dP_{1}(j,t)}{dt}=-(k_{-4}+k_{1}+k_{x})P_{1}(j,t)+k_{-1}P_{2}(j,t)+k_{p}P_{5}(j,t) 
\end{equation}
\begin{equation}
\dfrac{dP_{2}(j,t)}{dt}=k_{1}P_{1}(j,t)-(k_{-1}+k_{2})P_{2}(j,t)+k_{-2}P_{3}(j,t)
\end{equation}
\begin{equation}
\dfrac{dP_{3}(j,t)}{dt}=k_{2}P_{2}(j,t)-(k_{-2}+k_{3})P_{3}(j,t)+k_{-3}P_{4}(j,t)
\end{equation}
\begin{equation}
\dfrac{dP_{4}(j,t)}{dt}=k_{3}P_{3}(j,t)-(k_{4}+k_{-3})P_{4}(j,t)
\end{equation}
\begin{equation}
\dfrac{dP_{5}(j,t)}{dt}=k_{x}P_{1}(j,t)-(k_{p}+k_{exo})P_{5}(j,t)
\end{equation}
These equation can be re-expressed in the following matrix form.
\begin{equation}
\dfrac{d}{dt}{\bf P(t)}= {\bf M P(t)}
\label{main1}
\end{equation}
Here {\bf P(t)} is a column matrix, whose elements are $P_{1}(j,t)$, $P_{2}(j,t)$, 
$P_{3}(j,t)$, $P_{4}(j,t)$ and $P_{5}(j,t)$. And 
\begin{equation}
\textbf{M} =
\begin{bmatrix}
    -(k_{-4}+k_{1}+k_{x}) & k_{-1} & 0 & 0 &k_{p} \\
    k_{1} & -(k_{-1}+k_{2}) & k_{-2} & 0 & 0 \\
   0 & k_{2} & -(k_{-2}+k_{3}) & k_{-3} & 0 \\
   0 & 0 & k_{3} &-(k_{4}+k_{-3}) & 0 \\
   k_{x} & 0 & 0 & 0 & -(k_{p}+k_{exo}) \\ 
\end{bmatrix}
\end{equation}
Now introducing the Laplace transform of the probability of kinetic states by, 
\begin{equation}
\tilde{P_{\mu}}(j,s)=\int_{0}^{\infty}P_{\mu}(j,t)e^{-st}dt
\end{equation}
Solution of equation (\ref{main1}) in Laplace space is, 
\begin{equation}
{\bf \tilde{P}}(j,s)= (s{\bf I}-{\bf M})^{-1}{\bf {\tilde P}}(j,0)
\label{main2}
\end{equation}
Here ${\bf \tilde{P}}(j,s)$ is the vector of the probability of individual chemical 
state in Laplace 
space and ${\bf {\tilde P}}(j,0)$ is the column vector of initial probabilities.\\
Determinant of matrix $s{\bf I}-{\bf M}$ is a fifth order polynomial 
\begin{equation}
det(s{\bf I}-{\bf M})=\alpha s^5+\beta s^4+\gamma s^3+ \delta s^2 +\epsilon s + \zeta;
\end{equation}
full expressions for $\alpha$, $\beta$, $\gamma$, $\delta$, $\epsilon$ and $\zeta$ in terms of the primary rate constants are given in Appendix C. 

\subsubsection{Calculation of $\Psi_{++}, \Psi_{+-}, \Psi_{+x}$}

Following set of initial conditions guarantees that previous step taken by DNA polymerase
is a forward step.
\begin{equation}
P_{1}(j,0)=1 ~{\rm and}~ P_{2}(j,0)= P_{3}(j,0)= P_{4}(j,0)= P_{5}(j,0)=0 
\label{initial1}
\end{equation}
So three different distribution of dwell time, where first step is forward, are 
defined as follows:
\begin{equation}
\Psi_{++}(t)=P_{4}(j,t)k_{4}|_{[P_{1}(j,0)=1, P_{2}(j,0)= P_{3}(j,0)= P_{4}(j,0)= P_{5}(j,0)=0]}
\label{sipp}
\end{equation}
\begin{equation}
\Psi_{+-}(t)=P_{1}(j,t)k_{-4}|_{[P_{1}(j,0)=1, P_{2}(j,0)= P_{3}(j,0)= P_{4}(j,0)= P_{5}(j,0)=0]}
\label{sipn}
\end{equation}
\begin{equation}
\Psi_{+x}(t)=P_{5}(j,t)k_{exo}|_{[P_{1}(j,0)=1, P_{2}(j,0)= P_{3}(j,0)= P_{4}(j,0)= P_{5}(j,0)=0]}
\label{sipx}
\end{equation}
By applying the initial condition (\ref{initial1}) in equation (\ref{main2}), we get
\begin{equation}
\tilde{P_{4}}(j,s)=\dfrac{a_{0}+a_{1}s}{\alpha s^5+\beta s^4+\gamma s^3+\delta s^2+\epsilon s+\zeta}
\label{flpp}
\end{equation}
\begin{equation}
\tilde{P_{1}}(j,s)=\dfrac{b_{4}s^{4}+b_{3}s^{3}+b_{2}s^{2}+b_{1}s+b_{0}}{\alpha s^5+\beta s^4+\gamma s^3+\delta s^2+\epsilon s+\zeta}
\label{flpn}
\end{equation}
\begin{equation}
\tilde{P_{5}}(j,s)=\dfrac{c_{3}s^3+c_{2}s^{2}+c_{1}s+c_{0}}{\alpha s^5+\beta s^4+\gamma s^3+\delta s^2+\epsilon s+\zeta}
\label{flpx}
\end{equation}
Mathematical expressions for $a_{0}$, $a_{1}$, $b_{0}$, $b_{1}$, $b_{2}$, $b_{3}$, $b_{4}$,
$c_{0}$, $c_{1}$, $c_{2}$ and $c_{3}$ are given in Appendix D.\\
By inserting the inverse Laplace transforms of the expressions (\ref{flpp}), (\ref{flpn}) and (\ref{flpx})  into the equations (\ref{sipp}), (\ref{sipn}) and (\ref{sipx}), respectively, we get

\begin{figure}
\begin{center}
\includegraphics[angle=-90,width=0.85\columnwidth]{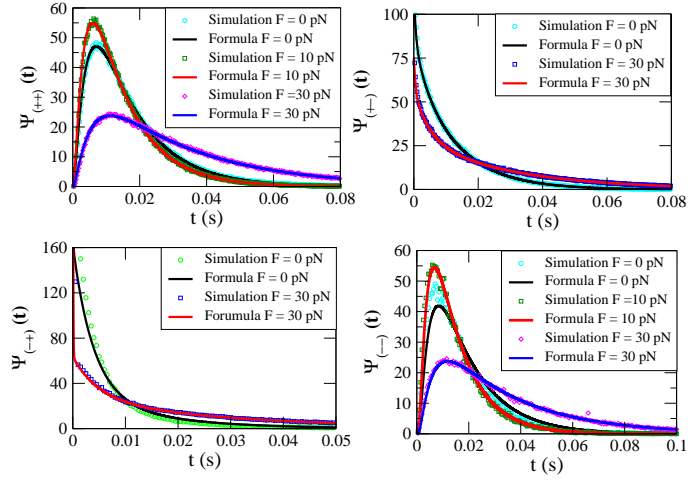}
\end{center}
\caption{$\Psi_{++}(t)$, $\Psi_{+-}(t)$, $\Psi_{-+}(t)$, and $\Psi_{--}(t)$ are plotted for a few different values of F.}
\label{f++}
\end{figure}

\begin{figure}
\begin{center}
\includegraphics[angle=-90,width=0.85\columnwidth]{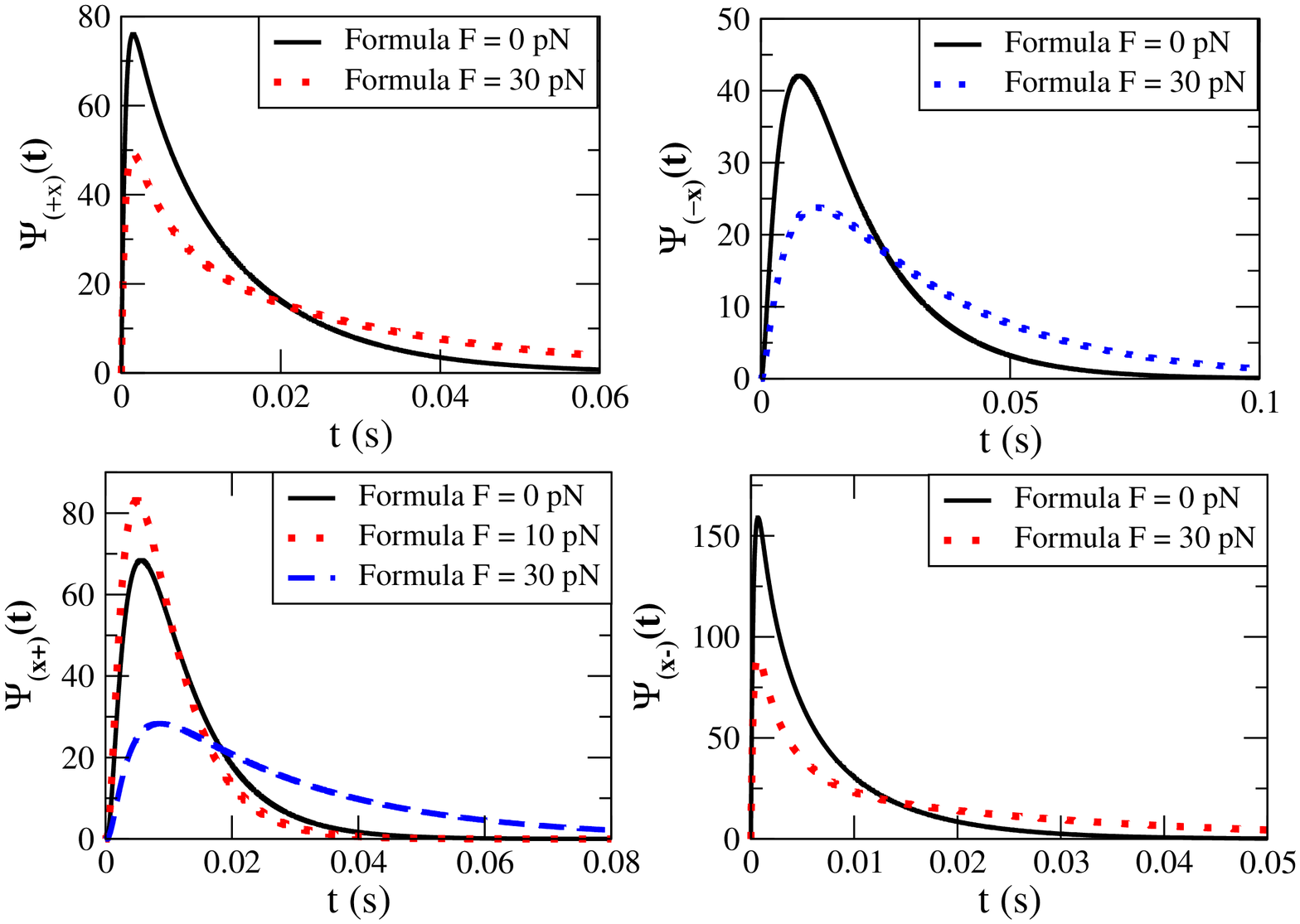}
\end{center}
\caption{$\Psi_{+x}(t)$, $\Psi_{-x}(t)$, $\Psi_{x+}(t)$, and $\Psi_{x-}(t)$ are plotted for a few different values of F.}
\label{fxx}
\end{figure}

\begin{eqnarray}
\Psi_{++}(t)&=&\biggl[\dfrac{(a_{0}-a_{1}\omega_{1})k_{4}}{(\omega_{1}-\omega_{2})(\omega_{1}-\omega_{3})(\omega_{1}-\omega_{4})(\omega_{1}-\omega_{5})}\biggr]e^{-\omega_{1}t} \nonumber \\ &+& \biggl[\dfrac{(a_{0}-a_{1}\omega_{2})k_{4}}{(\omega_{2}-\omega_{1})(\omega_{2}-\omega_{3})(\omega_{2}-\omega_{4})(\omega_{2}-\omega_{5})}\biggr]e^{-\omega_{2}t} \nonumber \\ &+& \biggl[\dfrac{(a_{0}-a_{1}\omega_{3})k_{4}}{(\omega_{3}-\omega_{1})(\omega_{3}-\omega_{2})(\omega_{3}-\omega_{4})(\omega_{3}-\omega_{5})}\biggr]e^{-\omega_{3}t} \nonumber \\ &+& \biggl[\dfrac{(a_{0}-a_{1}\omega_{4})k_{4}}{(\omega_{4}-\omega_{1})(\omega_{4}-\omega_{2})(\omega_{4}-\omega_{3})(\omega_{4}-\omega_{5})}\biggr]e^{-\omega_{4}t} \nonumber \\ &+& 
\biggl[\dfrac{(a_{0}-a_{1}\omega_{5})k_{4}}{(\omega_{5}-\omega_{1})(\omega_{5}-\omega_{2})(\omega_{5}-\omega_{3})(\omega_{5}-\omega_{4})}\biggr]e^{-\omega_{5}t}  
\end{eqnarray}
\begin{eqnarray}
\Psi_{+-}(t)&=&\biggl[\dfrac{(b_{0}-b_{1}\omega_{1}+b_{2}\omega_{1}^{2}-b_{3}\omega_{1}^{3}
+b_{4}\omega_{1}^{4})k_{-4}}{(\omega_{1}-\omega_{2})(\omega_{1}-\omega_{3})(\omega_{1}-\omega_{4})(\omega_{1}-\omega_{5})}\biggr]e^{-\omega_{1}t} \nonumber \\ &+& \biggl[\dfrac{(b_{0}-b_{1}\omega_{2}+b_{2}\omega_{2}^{2}-b_{3}\omega_{2}^{3}+b_{4}\omega_{2}^{4})k_{-4}}{(\omega_{2}-\omega_{1})(\omega_{2}-\omega_{3})(\omega_{2}-\omega_{4})(\omega_{2}-\omega_{5})}\biggr]e^{-\omega_{2}t} \nonumber \\ &+& \biggl[\dfrac{(b_{0}-b_{1}\omega_{3}+b_{2}\omega_{3}^{2}-b_{3}\omega_{3}^{3}+b_{4}\omega_{3}^{4})k_{-4}}{(\omega_{3}-\omega_{1})(\omega_{3}-\omega_{2})(\omega_{3}-\omega_{4})(\omega_{3}-\omega_{5})}\biggr]e^{-\omega_{3}t} \nonumber \\ &+& \biggl[\dfrac{(b_{0}-b_{1}\omega_{4}+b_{2}\omega_{4}^{2}-b_{3}\omega_{4}^{3}+b_{4}\omega_{4}^{4})k_{-4}}{(\omega_{4}-\omega_{1})(\omega_{4}-\omega_{2})(\omega_{4}-\omega_{3})(\omega_{4}-\omega_{5})}\biggr]e^{-\omega_{4}t} \nonumber \\ &+& 
\biggl[\dfrac{(b_{0}-b_{1}\omega_{5}+b_{2}\omega_{5}^{2}-b_{3}\omega_{5}^{3}+b_{4}\omega_{5}^{4})k_{-4}}{(\omega_{5}-\omega_{1})(\omega_{5}-\omega_{2})(\omega_{5}-\omega_{3})(\omega_{5}-\omega_{4})}\biggr]e^{-\omega_{5}t}  
\end{eqnarray}
\begin{eqnarray}
\Psi_{+x}(t)&=&\biggl[\dfrac{(c_{0}-c_{1}\omega_{1}+c_{2}\omega_{1}^{2}-c_{3}\omega_{1}^{3})k_{exo}}{(\omega_{1}-\omega_{2})(\omega_{1}-\omega_{3})(\omega_{1}-\omega_{4})(\omega_{1}-\omega_{5})}\biggr]e^{-\omega_{1}t} \nonumber \\ &+& \biggl[\dfrac{(c_{0}-c_{1}\omega_{2}+c_{2}\omega_{2}^{2}-c_{3}\omega_{2}^{3})k_{exo}}{(\omega_{2}-\omega_{1})(\omega_{2}-\omega_{3})(\omega_{2}-\omega_{4})(\omega_{2}-\omega_{5})}\biggr]e^{-\omega_{2}t} \nonumber \\ &+& \biggl[\dfrac{(c_{0}-c_{1}\omega_{3}+c_{2}\omega_{3}^{2}-c_{3}\omega_{3}^{3})k_{exo}}{(\omega_{3}-\omega_{1})(\omega_{3}-\omega_{2})(\omega_{3}-\omega_{4})(\omega_{3}-\omega_{5})}\biggr]e^{-\omega_{3}t} \nonumber \\ &+& \biggl[\dfrac{(c_{0}-c_{1}\omega_{4}+c_{2}\omega_{4}^{2}-c_{3}\omega_{4}^{3})k_{exo}}{(\omega_{4}-\omega_{1})(\omega_{4}-\omega_{2})(\omega_{4}-\omega_{3})(\omega_{4}-\omega_{5})}\biggr]e^{-\omega_{4}t} \nonumber \\ &+& 
\biggl[\dfrac{(c_{0}-c_{1}\omega_{5}+c_{2}\omega_{5}^{2}-c_{3}\omega_{5}^{3})k_{exo}}{(\omega_{5}-\omega_{1})(\omega_{5}-\omega_{2})(\omega_{5}-\omega_{3})(\omega_{5}-\omega_{4})}\biggr]e^{-\omega_{5}t}  
\end{eqnarray}
where $\omega_{1}$, $\omega_{2}$, $\omega_{3}$, $\omega_{4}$ and $\omega_{5}$ are the roots of
following equation
\begin{equation}
\alpha \omega^{5}-\beta \omega^{4}+\gamma \omega^{3}-\delta \omega_{2}+\epsilon \omega-\zeta=0;
\label{eq-omega5}
\end{equation} 
the explicit expressions of $\alpha, \beta, \gamma, \delta, \epsilon$ and $\zeta$ in terms of the primary rate constants of the kinetic model are given in appendix C.
The coupled nature of the pol and exo activities is revealed by the 
mixing of the corresponding rate constants in the expressions of 
$\Psi_{+,\pm}$ and $\Psi_{+x}$.

\subsubsection{Calculation of $\Psi_{-+}, \Psi_{--}, \Psi_{-x}$}

Following initial conditions ensures that DNA polymerase has reached to site $j$ by making
a backward step:
\begin{equation}
P_{4}(j,0)=1 ~{\rm and}~ P_{1}(j,0)= P_{2}(j,0)= P_{3}(j,0)= P_{5}(j,0)=0 
\end{equation}
So three different distributions of dwell time, where first step is backward, are 
defined as follows:
\begin{equation}
\Psi_{-+}(t)=P_{4}(j,t)k_{4}|_{[P_{4}(j,0)=1, P_{1}(j,0)= P_{2}(j,0)= P_{3}(j,0)= P_{5}(j,0)=0]}
\label{fsnp}
\end{equation}
\begin{equation}
\Psi_{--}(t)=P_{1}(j,t)k_{-4}|_{[P_{4}(j,0)=1, P_{1}(j,0)= P_{2}(j,0)= P_{3}(j,0)= P_{5}(j,0)=0]}
\label{fsnn}
\end{equation}
\begin{equation}
\Psi_{-x}(t)=P_{5}(j,t)k_{exo}|_{[P_{4}(j,0)=1, P_{1}(j,0)= P_{2}(j,0)= P_{3}(j,0)= P_{5}(j,0)=0]}
\label{fsnx}
\end{equation}
After applying the above initial condition in equation ({\ref{main2}}), we get
\begin{equation}
\tilde{P_{4}}(j,s)=\dfrac{d_{4}s^4+d_{3}s^3+d_{2}s^2+d_{1}s+d_{0}}{\alpha s^5+\beta s^4+\gamma s^3+\delta s^2+\epsilon s+\zeta}
\label{flnp}
\end{equation}
\begin{equation}
\tilde{P_{1}}(j,s)=\dfrac{k_{x}k_{-1}k_{-2}k_{-3}}{\alpha s^5+\beta s^4+\gamma s^3+\delta s^2+\epsilon s+\zeta}
\label{flnn}
\end{equation}
\begin{equation}
\tilde{P_{5}}(j,s)=\dfrac{e_{0}+e_{1}s}{\alpha s^5+\beta s^4+\gamma s^3+\delta s^2+\epsilon s+\zeta}
\label{flnx}
\end{equation}
Full expressions for $d_{0}$, $d_{1}$, $d_{2}$, $d_{3}$, $d_{4}$, $e_{0}$ and $e_{1}$ in terms of the primary rate constants of the kinetic model are given in Appendix D. Inverse transform of equation (\ref{flnp}), (\ref{flnn}) and (\ref{flnx})
gives the mathematical expression for $P_{4}(j,t)$, $P_{1}(j,t)$ and $P_{5}(j,t)$. 

Substituting the inverse Laplace transforms of (\ref{flnp}), (\ref{flnn}) and (\ref{flnx}) into 
the equations (\ref{fsnp}), (\ref{fsnn})
and (\ref{fsnx}) respectively, we get the following distributions of the conditional dwell time:  
\begin{eqnarray}
\Psi_{-+}(t)&=&\biggl[\dfrac{(d_{0}-d_{1}\omega_{1}+d_{2}\omega_{1}^{2}-d_{3}\omega_{1}^{3}+d_{4}\omega_{1}^{4})k_{4}}{(\omega_{1}-\omega_{2})(\omega_{1}-\omega_{3})(\omega_{1}-\omega_{4})(\omega_{1}-\omega_{5})}\biggr]e^{-\omega_{1}t} \nonumber \\ &+& \biggl[\dfrac{(d_{0}-d_{1}\omega_{2}+d_{2}\omega_{2}^{2}-d_{3}\omega_{2}^{3}+d_{4}\omega_{2}^{4})k_{4}}{(\omega_{2}-\omega_{1})(\omega_{2}-\omega_{3})(\omega_{2}-\omega_{4})(\omega_{2}-\omega_{5})}\biggr]e^{-\omega_{2}t} \nonumber \\ &+& \biggl[\dfrac{(d_{0}-d_{1}\omega_{3}+d_{2}\omega_{3}^{2}-d_{3}\omega_{3}^{3}+d_{4}\omega_{3}^{4})k_{4}}{(\omega_{3}-\omega_{1})(\omega_{3}-\omega_{2})(\omega_{3}-\omega_{4})(\omega_{3}-\omega_{5})}\biggr]e^{-\omega_{3}t} \nonumber \\ &+& \biggl[\dfrac{(d_{0}-d_{1}\omega_{4}+d_{2}\omega_{4}^{2}-d_{3}\omega_{4}^{3}+d_{4}\omega_{4}^{4})k_{4}}{(\omega_{4}-\omega_{1})(\omega_{4}-\omega_{2})(\omega_{4}-\omega_{3})(\omega_{4}-\omega_{5})}\biggr]e^{-\omega_{4}t} \nonumber \\ &+& 
\biggl[\dfrac{(d_{0}-d_{1}\omega_{5}+d_{2}\omega_{5}^{2}-d_{3}\omega_{5}^{3}+d_{4}\omega_{5}^{4})k_{4}}{(\omega_{5}-\omega_{1})(\omega_{5}-\omega_{2})(\omega_{5}-\omega_{3})(\omega_{5}-\omega_{4})}\biggr]e^{-\omega_{5}t}  
\end{eqnarray}

\begin{eqnarray}
\Psi_{--}(t)&=&\biggl[\dfrac{k_{x}k_{-1}k_{-2}k_{-3}k_{-4}}{(\omega_{1}-\omega_{2})(\omega_{1}-\omega_{3})(\omega_{1}-\omega_{4})(\omega_{1}-\omega_{5})}\biggr]e^{-\omega_{1}t} \nonumber \\ &+& \biggl[\dfrac{k_{x}k_{-1}k_{-2}k_{-3}k_{-4}}{(\omega_{2}-\omega_{1})(\omega_{2}-\omega_{3})(\omega_{2}-\omega_{4})(\omega_{2}-\omega_{5})}\biggr]e^{-\omega_{2}t} \nonumber \\ &+& \biggl[\dfrac{k_{x}k_{-1}k_{-2}k_{-3}k_{-4}}{(\omega_{3}-\omega_{1})(\omega_{3}-\omega_{2})(\omega_{3}-\omega_{4})(\omega_{3}-\omega_{5})}\biggr]e^{-\omega_{3}t} \nonumber \\ &+& \biggl[\dfrac{k_{x}k_{-1}k_{-2}k_{-3}k_{-4}}{(\omega_{4}-\omega_{1})(\omega_{4}-\omega_{2})(\omega_{4}-\omega_{3})(\omega_{4}-\omega_{5})}\biggr]e^{-\omega_{4}t} \nonumber \\ &+& 
\biggl[\dfrac{k_{x}k_{-1}k_{-2}k_{-3}k_{-4}}{(\omega_{5}-\omega_{1})(\omega_{5}-\omega_{2})(\omega_{5}-\omega_{3})(\omega_{5}-\omega_{4})}\biggr]e^{-\omega_{5}t}  
\end{eqnarray}

\begin{eqnarray}
\Psi_{-x}(t)&=&\biggl[\dfrac{(e_{0}-e_{1}\omega_{1})k_{exo}}{(\omega_{1}-\omega_{2})(\omega_{1}-\omega_{3})(\omega_{1}-\omega_{4})(\omega_{1}-\omega_{5})}\biggr]e^{-\omega_{1}t} \nonumber \\ &+& \biggl[\dfrac{(e_{0}-e_{1}\omega_{2})k_{exo}}{(\omega_{2}-\omega_{1})(\omega_{2}-\omega_{3})(\omega_{2}-\omega_{4})(\omega_{2}-\omega_{5})}\biggr]e^{-\omega_{2}t} \nonumber \\ &+& \biggl[\dfrac{(e_{0}-e_{1}\omega_{3})k_{exo}}{(\omega_{3}-\omega_{1})(\omega_{3}-\omega_{2})(\omega_{3}-\omega_{4})(\omega_{3}-\omega_{5})}\biggr]e^{-\omega_{3}t} \nonumber \\ &+& \biggl[\dfrac{(e_{0}-e_{1}\omega_{4})k_{exo}}{(\omega_{4}-\omega_{1})(\omega_{4}-\omega_{2})(\omega_{4}-\omega_{3})(\omega_{4}-\omega_{5})}\biggr]e^{-\omega_{4}t} \nonumber \\ &+& 
\biggl[\dfrac{(e_{0}-e_{1}\omega_{5})k_{exo}}{(\omega_{5}-\omega_{1})(\omega_{5}-\omega_{2})(\omega_{5}-\omega_{3})(\omega_{5}-\omega_{4})}\biggr]e^{-\omega_{5}t}  
\end{eqnarray}
where $\omega_{1}$, $\omega_{2}$, $\omega_{3}$, $\omega_{4}$ and $\omega_{5}$ are the roots of
the equation (\ref{eq-omega5}).

\subsubsection{Calculation of $\Psi_{x+}, \Psi_{x-}, \Psi_{xx}$}

Now we consider the case where DNA polymerase has arrived at site $i$ by making
an exonuclease cleavage. The initial condition   
\begin{equation}
P_{5}(j,0)=1 ~{\rm and}~ P_{1}(j,0)= P_{2}(j,0)= P_{3}(j,0)= P_{4}(j,0)=0 
\end{equation}
ensures that previous mechanical step is an exonuclease cleaving. Now we define
following distributions of conditional dwell time  
\begin{equation}
\Psi_{x+}(t)=P_{4}(j,t)k_{4}|_{[P_{5}(j,0)=1, P_{1}(j,0)= P_{2}(j,0)= P_{3}(j,0)= P_{4}(j,0)=0]}
\label{fsxp}
\end{equation}
\begin{equation}
\Psi_{x-}(t)=P_{1}(j,t)k_{-4}|_{[P_{5}(j,0)=1, P_{1}(j,0)= P_{2}(j,0)= P_{3}(j,0)= P_{4}(j,0)=0]}
\label{fsxn}
\end{equation}
\begin{equation}
\Psi_{xx}(t)=P_{5}(j,t)k_{exo}|_{[P_{5}(j,0)=1, P_{1}(j,0)= P_{2}(j,0)= P_{3}(j,0)= P_{4}(j,0)=0]}
\label{fsxx}
\end{equation}
After applying the above initial condition in equation \ref{main2}, we get
\begin{equation}
\tilde{P_{5}}(j,s)=\dfrac{f_{4}s^4+f_{3}s^3+f_{2}s^2+f_{1}s+f_{0}}{\alpha s^5 + \beta s^4+\gamma s^3+\delta s^2+\gamma s
+\zeta}
\label{flxp}
\end{equation}
\begin{equation}
\tilde{P_{4}}(j,s)=\dfrac{k_{1}k_{2}k_{3}k_{p}}{\alpha s^5+\beta s^4+\gamma s^3+\delta s^2+\epsilon s+\zeta}
\label{flxn}
\end{equation}
\begin{equation}
\tilde{P_{1}}(j,s)=\dfrac{g_{3}s^3+g_{2}s^2+g_{1}s+g_{0}}{\alpha s^5+\beta s^4+\gamma s^3+\delta s^2+\epsilon s+\zeta}
\label{flxx}
\end{equation}
The expressions for $f_{0}$, $f_{1}$, $f_{2}$, $f_{3}$, $f_{4}$, $g_{0}$, $g_{1}$, $g_{2}$
and $g_{3}$ are given in Appendix D. The values of $P_{4}(j,t)$, $P_{1}(j,t)$ and 
$P_{5}(j,t)$ are obtained from the inverse Laplace transform of the (\ref{flxp}), (\ref{flxn}) 
and (\ref{flxn}). After inserting the values of $P_{4}(j,t)$, $P_{1}(j,t)$ and $P_{5}(j,t)$
in equations (\ref{fsxp}), (\ref{fsxn}) and (\ref{fsxn}), we get the exact
analytical expression for $\Psi{xx}(t)$, $\Psi{x+}(t)$ and $\Psi{x-}(t)$.

\begin{eqnarray}
\Psi_{xx}(t)&=&\biggl[\dfrac{(f_{0}-f_{1}\omega_{1}+f_{2}\omega_{1}^{2}-f_{3}\omega_{1}^{3}+f_{4}\omega_{1}^{4})k_{exo}}{(\omega_{1}-\omega_{2})(\omega_{1}-\omega_{3})(\omega_{1}-\omega_{4})(\omega_{1}-\omega_{5})}\biggr]e^{-\omega_{1}t} \nonumber \\ &+& \biggl[\dfrac{(f_{0}-f_{1}\omega_{2}+f_{2}\omega_{2}^{2}-f_{3}\omega_{2}^{3}+f_{4}\omega_{2}^{4})k_{exo}}{(\omega_{2}-\omega_{1})(\omega_{2}-\omega_{3})(\omega_{2}-\omega_{4})(\omega_{2}-\omega_{5})}\biggr]e^{-\omega_{2}t} \nonumber \\ &+& \biggl[\dfrac{(f_{0}-f_{1}\omega_{3}+f_{2}\omega_{3}^{2}-f_{3}\omega_{3}^{3}+f_{4}\omega_{3}^{4})k_{exo}}{(\omega_{3}-\omega_{1})(\omega_{3}-\omega_{2})(\omega_{3}-\omega_{4})(\omega_{3}-\omega_{5})}\biggr]e^{-\omega_{3}t} \nonumber \\ &+& \biggl[\dfrac{(f_{0}-f_{1}\omega_{4}+f_{2}\omega_{4}^{2}-f_{3}\omega_{4}^{3}+f_{4}\omega_{4}^{4})k_{exo}}{(\omega_{4}-\omega_{1})(\omega_{4}-\omega_{2})(\omega_{4}-\omega_{3})(\omega_{4}-\omega_{5})}\biggr]e^{-\omega_{4}t} \nonumber \\ &+& 
\biggl[\dfrac{(f_{0}-f_{1}\omega_{5}+f_{2}\omega_{5}^{2}-f_{3}\omega_{5}^{3}+f_{4}\omega_{5}^{4})k_{exo}}{(\omega_{5}-\omega_{1})(\omega_{5}-\omega_{2})(\omega_{5}-\omega_{3})(\omega_{5}-\omega_{4})}\biggr]e^{-\omega_{5}t}  
\end{eqnarray}
\begin{eqnarray}
\Psi_{x+}(t)&=&\biggl[\dfrac{k_{1}k_{2}k_{3}k_{4}k_{p}}{(\omega_{1}-\omega_{2})(\omega_{1}-\omega_{3})(\omega_{1}-\omega_{4})(\omega_{1}-\omega_{5})}\biggr]e^{-\omega_{1}t} \nonumber \\ &+& \biggl[\dfrac{k_{1}k_{2}k_{3}k_{4}k_{p}}{(\omega_{2}-\omega_{1})(\omega_{2}-\omega_{3})(\omega_{2}-\omega_{4})(\omega_{2}-\omega_{5})}\biggr]e^{-\omega_{2}t} \nonumber \\ &+& \biggl[\dfrac{k_{1}k_{2}k_{3}k_{4}k_{p}}{(\omega_{3}-\omega_{1})(\omega_{3}-\omega_{2})(\omega_{3}-\omega_{4})(\omega_{3}-\omega_{5})}\biggr]e^{-\omega_{3}t} \nonumber \\ &+& \biggl[\dfrac{k_{1}k_{2}k_{3}k_{4}k_{p}}{(\omega_{4}-\omega_{1})(\omega_{4}-\omega_{2})(\omega_{4}-\omega_{3})(\omega_{4}-\omega_{5})}\biggr]e^{-\omega_{4}t} \nonumber \\ &+& 
\biggl[\dfrac{k_{1}k_{2}k_{3}k_{4}k_{p}}{(\omega_{5}-\omega_{1})(\omega_{5}-\omega_{2})(\omega_{5}-\omega_{3})(\omega_{5}-\omega_{4})}\biggr]e^{-\omega_{5}t}  
\end{eqnarray}
\begin{eqnarray}
\Psi_{x-}(t)&=&\biggl[\dfrac{(g_{0}-g_{1}\omega_{1}+g_{2}\omega_{1}^{2}-g_{3}\omega_{1}^{3})k_{-4}}{(\omega_{1}-\omega_{2})(\omega_{1}-\omega_{3})(\omega_{1}-\omega_{4})(\omega_{1}-\omega_{5})}\biggr]e^{-\omega_{1}t} \nonumber \\ &+& \biggl[\dfrac{(g_{0}-g_{1}\omega_{2}+g_{2}\omega_{2}^{2}-g_{3}\omega_{2}^{3})k_{-4}}{(\omega_{2}-\omega_{1})(\omega_{2}-\omega_{3})(\omega_{2}-\omega_{4})(\omega_{2}-\omega_{5})}\biggr]e^{-\omega_{2}t} \nonumber \\ &+& \biggl[\dfrac{(g_{0}-g_{1}\omega_{3}+g_{2}\omega_{3}^{2}-g_{3}\omega_{3}^{3})k_{-4}}{(\omega_{3}-\omega_{1})(\omega_{3}-\omega_{2})(\omega_{3}-\omega_{4})(\omega_{3}-\omega_{5})}\biggr]e^{-\omega_{3}t} \nonumber \\ &+& \biggl[\dfrac{(g_{0}-g_{1}\omega_{4}+g_{2}\omega_{4}^{2}-g_{3}\omega_{4}^{3})k_{-4}}{(\omega_{4}-\omega_{1})(\omega_{4}-\omega_{2})(\omega_{4}-\omega_{3})(\omega_{4}-\omega_{5})}\biggr]e^{-\omega_{4}t} \nonumber \\ &+& 
\biggl[\dfrac{(g_{0}-g_{1}\omega_{5}+g_{2}\omega_{5}^{2}-g_{3}\omega_{5}^{3})k_{-4}}{(\omega_{5}-\omega_{1})(\omega_{5}-\omega_{2})(\omega_{5}-\omega_{3})(\omega_{5}-\omega_{4})}\biggr]e^{-\omega_{5}t}  
\end{eqnarray}
where $\omega_{1}$, $\omega_{2}$, $\omega_{3}$, $\omega_{4}$ and $\omega_{5}$ are the roots of
the equation (\ref{eq-omega5}).

The distributions of the conditional dwell times $\Psi_{mn}$, except 
$\Psi_{xx}$, are plotted for a few typical values of the parameters 
in figs.\ref{f++} and \ref{fxx}. Since $\Psi_{xx}$ is independent of 
the tension $F$, it has not been drawn graphically. We have also 
presented our numerical data, obtained from direct computer simulation, 
for the distributions plotted in fig.\ref{f++}.
Each of these distributions is a sum of several exponentials. Therefore, 
in general, these distributions are expected to peak at a nonzero 
value of time $t$. However, some of the distributions in fig.\ref{f++} 
and \ref{fxx} appear as a single exponential. This single-exponential 
like appearance is an artefact of the parameters chosen for plotting 
these curves although, in reality, the full distributions are sum of 
several exponentials. 

An interesting feature of the distributions plotted in figs.\ref{f++} and 
\ref{fxx} is a non-monotonic variation of the probability of the most 
probable conditional dwell times with increasing $F$ 
(see, for example, $\Psi_{++}$ and $\Psi_{--}$). This trend of variation 
is a consequence of the nonmonotonic variation of $\Delta \Phi$ with $F$ 
(see fig.\ref{phie}).

\section{Proposals for experimental test of the theoretical predictions}

The distributions of the conditional dwell times $\psi_{\pm\pm}$ have 
been extracted for some motors in the last decade from the data 
obtained from single-molecule experiments.  
But, to our knowledge, none of the distributions  
$\Psi_{\pm\pm}$, $\Psi_{x\pm}$, $\Psi_{\pm x}$ and $\Psi_{xx}$
have been measured experimentally so far specifically for the DNAP motor. 
In this section we first mention a few recently developed single-molecule 
techniques that probe some aspects of DNAP kinetics during replication. 

In a landmark paper Eid et al. \cite{eid09} reported a single-molecule 
method for monitoring replication exploiting fluorescently labelled 
nucleotide monomers. The fluorophores are ``Phospholinked'' (i.e., linked 
to the phosphate group of the nucleotide monomer) \cite{eid09}. Since 
DNAP-catalyzed phosphodiester bond formation releases the fluorophore from 
the nucleotide, the temporal sequence of the color of the fluorescence 
provides the sequence of the nucleotides that are incorporated in the 
elongating DNA. 
Christian et al.\cite{christian09} have developed a single-molecule 
technique for monitoring replication by a DNAP with base-pair resolution. 
This method is based on F\"orster resonance energy transfer (FRET). 
Use of this technique also makes it possible to discriminate between 
the polymerization activity and exonuclease activity of the DNAP.  
It is likely that in near future appropriate adaptations of these 
or some combination of force-based and fluorescence-based single molecule 
techniques may achieve sufficiently high resolution required for 
measuring the nine distributions of conditional dwell times introduced 
in this paper. 

Next, we propose a reduced description of the stochastic 
pause-and-translocation of the DNAP in terms of fewer conditional 
dwell times which, as we explain below, may be measurable with the 
currently available single molecule techniques because these do not 
distinguish between chemical and mechanical backward steppings. 
Let us define 
\begin{equation}
\Psi_{+}(t)=\Psi_{++}(t)+\Psi_{+-}(t)+\Psi_{+x}(t), 
\end{equation}
\begin{equation}
\Psi_{-}(t)=\Psi_{-+}(t)+\Psi_{--}(t)+\Psi_{-x}(t) 
\end{equation}
and
\begin{equation}
\Psi_{x}(t)=\Psi_{x+}(t)+\Psi_{x-}(t)+\Psi_{xx}(t) 
\end{equation}
as the distributions of conditional dwell times in which, regardless of the nature of the next step, the step taken by the DNAP is a forward polymerase step $(+)$, backward polymerase step $(-)$ and backward exonuclease step $(x)$, respectively.

For a given set of initial conditions, overall probability to leave the 
$j$th site should be unity. Therefore, following conditions must be 
satisfied:
\begin{equation}
\int_{0}^{\infty}(k_{-4}P_{1}(t)+k_{4}P_{4}(t)+k_{exo}P_{5}(t))|_{[P_{1}(j,0)=1 ~{\rm and}~ P_{i}(j,0)=0]}dt=\int_{0}^{\infty} \Psi_{+}(t)dt=1
\end{equation}
\begin{equation}
\int_{0}^{\infty}(k_{-4}P_{1}(t)+k_{4}P_{4}(t)+k_{exo}P_{5}(t))|_{[P_{4}(j,0)=1 ~{\rm and}~ P_{i}(j,0)=0]}dt=\int_{0}^{\infty} \Psi_{-}(t)dt=1
\end{equation}  
\begin{equation}
\int_{0}^{\infty}(k_{-4}P_{1}(t)+k_{4}P_{4}(t)+k_{exo}P_{5}(t))|_{[P_{5}(j,0)=1 ~{\rm and}~ P_{i}(j,0)=0]}dt=\int_{0}^{\infty} \Psi_{x}(t)dt=1
\end{equation} 
i.e., $\Psi_{+}(t)$, $\Psi_{-}(t)$ and $\Psi_{x}(t)$ are probability 
distributions normalized to unity.
Therefore, the overall distribution of dwell time, irrespective of the type of steps taken by DNAP, is the weighted sum
\begin{equation}
\Psi(t)=q_{+}\Psi_{+}(t)+q_{-}\Psi_{-}(t)+q_{x}\Psi_{x}(t).
\label{eq-wtuncond}
\end{equation} 
where $q_{+}$, $q_{-}$ and $q_{x}$ denote the probabilities of taking forward polymerase step $(+)$, backward polymerase step $(-)$ and backward exonuclease step $(x)$ by a DNAP, respectively. 
The explicit expressions for $q_{+}$, $q_{-}$ and $q_{x}$ are given by 
\begin{equation}
q_{+}=\dfrac{P_{4}^{st}k_{4}}{P_{4}^{st}k_{4}+P_{1}^{st}k_{-4}+P_{5}^{st}k_{exo}}=\dfrac{x_{4}k_{4}}{x_{4}k_{4}+x_{1}k_{-4}+x_{5}k_{exo}}
\label{eq-q+}
\end{equation}
\begin{equation}
q_{-}=\dfrac{P_{1}^{st}k_{-4}}{P_{4}^{st}k_{4}+P_{1}^{st}k_{-4}+P_{5}^{st}k_{exo}}=\dfrac{x_{1}k_{-4}}{x_{4}k_{4}+x_{1}k_{-4}+x_{5}k_{exo}}
\label{eq-q-}
\end{equation}
\begin{equation}
q_{x}=\dfrac{P_{5}^{st}k_{exo}}{P_{4}^{st}k_{4}+P_{1}^{st}k_{-4}+P_{5}^{st}k_{exo}}=\dfrac{x_{5}k_{exo}}{x_{4}k_{4}+x_{1}k_{-4}+x_{5}k_{exo}}
\label{eq-qx}
\end{equation}
where $x_{\mu}$s, in terms of the rate constants, are given in Appendix B.\\

We now recast eqn (\ref{eq-wtuncond}) in a form that would facilitate 
direct contact with experiments that are feasible with the currently 
available techniques. Writing 
\begin{equation}
\Psi(t)= \Xi_{++}(t)+\Xi_{+-}(t)+\Xi_{-+}(t)+\Xi_{--}(t)   
\end{equation}
we identify the four new distributions of conditional dwell times 
$\Xi_{\pm \pm}(t)$ to be 
\begin{equation}
\Xi_{++}(t)=q_{+}\Psi_{++}(t) 
\label{eq-Xi++}
\end{equation}
\begin{equation}
\Xi_{+-}(t)=q_{+}[\Psi_{+-}(t)+\Psi_{+x}(t)]
\label{eq-Xi+-}
\end{equation}
\begin{equation}
\Xi_{-+}(t)=q_{-}\Psi_{-+}(t)+q_{x}\Psi_{x+}(t)
\label{eq-Xi-+}
\end{equation}
\begin{equation}
\Xi_{--}(t)=q_{-}[\Psi_{--}(t)+\Psi_{-x}(t)]+q_{x}[\Psi_{x-}(t)+\Psi_{xx}(t)]
\label{eq-Xi--}
\end{equation}
where the symbols "$+$" and "$-$" denote forward and backward movements 
of the DNAP irrespective of the mode of movement. For example, the DNAP 
can move backward either by polymerase or exonuclease activity; however, 
the newly defined conditional dwell times $\Xi_{-}(t)$ does not 
discriminate between these two modes of backward movement. For the purpose 
of comparison with experimental data, expressions (\ref{eq-q+}),(\ref{eq-q-}) 
and (\ref{eq-qx}) for $q_{+}, q_{-}$ and $q_{x}$ and the expressions 
derived in section \ref{sec-results} for the conditional dwell times 
$\Psi_{\pm\pm}, \Psi_{\pm x}, \Psi_{x\pm}$ should be substituted into 
the eqns.(\ref{eq-Xi++})-(\ref{eq-Xi--}).

For a DNAP with the data set given in table 1, the probabilities 
for a polymerase-dependent forward step ($+$), polymerase-dependent 
backward step ($-$) and exonuclease activity ($x$) are 0.9736, .0261 
and .0003, respectively. Thus, under normal circumstances back-stepping 
and exonuclease events are very unlikely. Moreover, two consecutive {\it 
exonuclease} steps would be extremely rare. However, the frequency of {\it 
exonuclease} activity of the DNAP can be increased by using mutants of 
the same DNAP. By increasing the concentration of pyrophosphate far 
above the equilibrium concentration, back-stepping events can be made 
more frequent \cite{goel03}. Besides, transfer-deficient mutants and 
exonuclease-deficient mutants \cite{ibarra09} can be used to test the 
effects of variation of the corresponding rate constants on the various 
dwell time distributions.

\section{Summary and conclusion}

DNA replication is carried out by DNAP which operates as a molecular motor 
utilizing the template DNA strand as its track. In this paper we have 
presented a theoretical model for DNA replication that allows systematic 
investigation of the pol and exo activities as well as their coupling. 
More specifically, the situation considered here mimics an {\it in-vitro} 
experiment where a tension is applied on the template strand throughout 
the replication process. We have calculated the effect of the tension on 
the average speed of replication, capturing the effects of both the pol 
and exo activities of the same DNAP. Our theoretical results in section 
\ref{sec-fvcurve} are in good qualitative agreement with the results of 
single molecule experiments reported in the literature \cite{wuite00,goel03}. 

However, the intrinsic fluctuations in the pol and exo processes contain 
some additional information which cannot be extracted from average 
properties. It is well known that the fluctuation in the dwell times 
provides a numerical estimate the number of kinetic states 
\cite{moffitt10b,schnitzer95}. 
More specifically, one defines a ``randomness parameter'' 
$r = (<\tau^2>-<\tau>^2)/<\tau>^{2}$ where $\tau$ is the dwell time 
and the symbol $<.>$ indicates average; $1/r$ provides a lower bound 
on the number of kinetic states in each mechano-chemical cycle of the 
motor. Moreover, if $r$ is larger than unity in any parameter regime, 
it would indicate existence of branched pathways. Furthermore, 
{\it conditional} dwell times can reveal existence of correlations 
between individual steps of the mechano-chemical pathways of a 
molecular motor \cite{moffitt10b}. Besides, hidden substeps may be 
missed in the noisy data recorded in a single-motor experiment; the 
distributions of conditional dwell times are quite useful in detecting 
such substeps. Exact analytical expressions for the distributions of 
conditional dwell times that we report here may find use in the analysis 
of the experimental data for extracting these information \cite{tsygankov07}.

Although both the pol and exo activities of the DNAP have been studied 
extensively \cite{krantz10}, the distributions of dwell times of DNAP 
have not been measured so far in any single molecule experiment 
\cite{manosas12}. In this paper we have also 
mentioned a few recently developed single-molecule techniques for DNAP 
\cite{eid09,christian09} which, after minor alteration, might be the 
appropriate tool for measuring the conditional dwell times introduced in 
this paper. We have also proposed a reduced description of the 
pause-and-translocation of DNAP in terms of the distributions of fewer 
conditional dwell times which, in principle, can be measured by the 
currently available single-molecular techniques.
We hope our model and results will motivate experiments to 
study the unexplored stochastic features of the kinetics of one of the 
most important genetic processes, namely DNA replication driven by DNAP. 
Understanding this kinetics will throw light on the propagation of life 
from one generation to the next. \\

\noindent{\bf Acknowledgements}: DC thanks Berenike Maier for useful discussions. We also thank the anonymous referees for constructive criticism and suggestions which helped in significant improvement of the manuscript. This work has been supported by the Dr. Jagmohan Garg Chair Professorship (DC), J.C. Bose national fellowship (DC), Department of Biotechnology (DC) and Council of Scientific and Industrial Research (AKS). 

\noindent{\bf Appendix A}

Here the parameters with subscripts ``1'' and ``2'' correspond to ssDNA and dsDNA, respectively. Let $b_{i}(F)$ ($i=1,2$) denote the average equilibrium projections of base pair in the direction of the applied force $F$. Suppose, $\Phi_{i}(F)$ ($i=1,2)$ are the corresponding free energies. Then, for a given force $F$, the free energy difference between single base-pair of dsDNA and ssDNA can be expressed as \cite{rouzina01} 
\begin{equation}
\Delta \Phi'(F)=\Phi_{2}(F)-\Phi_{1}(F)=-\int_{0}^{F}(b_{2}(F')-b_{1}(F'))dF'
\label{eq-fdiff}
\end{equation}
where the right-hand side can be evaluated if the functions $b_{i}(F)$ are known.

For the freely jointed chain (FJC) model of DNA, 
is established.
\begin{equation}
b_{i}(F)=\biggl[coth\biggl(\dfrac{2FA_{i}}{k_{B}T}\biggr)-\dfrac{k_{B}T}{2FA_{i}}\biggr]\biggl(1+\dfrac{F}{K_{i}}\biggl)b_{i}^{max}
\label{eq-FJC}
\end{equation}
where $K_{i}$, $A_{i}$ and $b_{i}^{max}$ are, respectively, the elastic modulus, the persistence length and the average length of a base pair in the absence of any force. 

\begin{figure}
\begin{center}
\includegraphics[angle=-90,width=0.75\columnwidth]{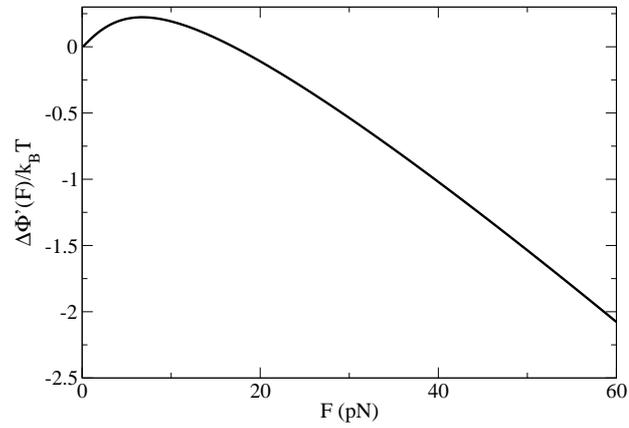}
\end{center}
\caption{Free energy difference $\Delta \Phi'$ is plotted against the
tension $F$.}
\label{phie}
\end{figure}

Inserting the expression (\ref{eq-FJC}) into the equation (\ref{eq-fdiff}) we numerically compute the free energy difference between single base pair of dsDNA and that of ssDNA for the given force $F$.  In figure \ref{phie} we plot $\Delta \Phi'$ against the tension $F$. The numerical values of the parameters that we use for this computation are given in the table \ref{tab-fdiff}. 

\begin{table}
\begin{tabular}{|l|l|}
  \hline
  \multicolumn{2}{|c|}{Parameter values}\\ 
  \hline
  $b_{1}^{max}$ & .58 nm \\ \hline
  $b_{2}^{max}$ & .34 nm \\  \hline
  $A_{1}$ & .7 nm \\  \hline
  $A_{2}$ &  50 nm \\  \hline
  $K_{1}$ &  900 pN \\  \hline
  $K_{2}$ &  1000 pN  \\  \hline
  \end{tabular} 
\caption{Numerical values of the relevant parameters used for the computation of $\Delta \Phi'$ using equation (\ref{eq-fdiff}) and (\ref{eq-FJC})}.
\label{tab-fdiff}
\end{table}

We now assume that change in the barrier height $\Delta \Phi (F)$ 
that enters into the equation (\ref{eq-fdeprate}) is equivalent 
to $n \Delta \Phi'(F)$ where $n>1$ is an integer. We would like to 
emphasize that $\Delta \Phi'(F)$ is the stetching free energy difference 
between ssDNA and dsDNA i.e., between the initial state 3 and final 
state 4 of the transition for which the barrier height, i.e., the 
free energy difference between state 3 and the transition state, 
contains the force-induced extra term $\Delta \Phi(F)$. Our 
assumption $\Delta \Phi(F) = n \Delta \Phi'(F) ~(n>1)$ is similar, 
in spirit, but not identical to the assumptions made by Wuite et al.  
\cite{wuite00} and Maier et al.\cite{maier00}. The physical meaning of 
this assumption is that the tension-induced change $\Delta \Phi(F)$ of 
the barrier arising from the legnth mismatch between the ssDNA and 
dsDNA base pairs is equivalent to $n$ times the free energy difference 
$\Delta \Phi'(F)$. Atomistic explanation of the physical origin of the 
tension-induced change $n \Delta \Phi'(F)$ of the activation energy 
would require more fine-grained modeling of the local neighborhood of 
the catalytic site \cite{andricioaei04} 
which is beyond the scope of the Markov kinetic models of the type 
developed in this paper. In our numerical calculations we use $n= 3$ 
which is consistent with the best fit values reported in 
refs.\cite{wuite00,maier00}. The parameter value $n=3$ should not be 
confused with the step size of the DNAP which is 1 nucleotide. 

The polymerase and exonuclease catalytic sites are separated by about 
3.5 nm. A DNA molecule migrating from the polymerase site to the 
exonuclease site of DNAP would cause local melting of more than one 
termial base pairs \cite{goel03,beese93,wang97,shamoo99}. Therefore, 
based on arguments similar to those used earlier for the rate 
constant $k_{3}(F)$, we now expect $\Delta \Phi _{x}(F) = m \Delta \Phi'(F)$. 
Since no further information is available to fix the numerical value of 
$m$, we use $m=3$ because this choice provides the best fit to the  
experimenta data \cite{wuite00}.

WLC model provides a slightly better quantitative estimate of the force 
extension curve of dsDNA in the range of 0 to 10 pN force 
\cite{bustamante10}. However, given the uncertainties of the other 
parameters used for plotting our results graphically, the simpler FJC 
model is good enough. Indeed, it produces the non monotonicity of 
$\Delta \Phi(F)$ as a function of force (F) which, in turn, can be used to 
estimate the mean rate of elongation as well as the conditonal dwell times.

\noindent{\bf Appendix B}

\begin{equation}
x_{1}=1
\end{equation}
\begin{equation}
x_{2}=\dfrac{k_{1}+x_{3}k_{-2}}{k_{-1}+k_{2}}
\end{equation}
\begin{equation}
x_{3}=\dfrac{k_{2}k_{1}(k_{4}+k_{-3})+k_{-3}k_{-4}(k_{-1}+k_{2})}{(k_{4}+k_{-3})(k_{-1}+k_{2})(k_{-2}+k_{3})-k_{-3}k_{3}(k_{-1}+k_{2})-k_{2}k_{-2}(k_{4}+k_{-3})}
\end{equation}
\begin{equation}
x_{4}=\dfrac{k_{-4}+x_{3}k_{3}}{k_{4}+k_{-3}}
\end{equation}
\begin{equation}
x_{5}=\dfrac{k_{x}}{k_{p}}
\end{equation}

\noindent{\bf Appendix C}

\begin{equation}
\alpha=1
\end{equation}
\begin{equation}
\beta = k_{1}+k_{2}+k_{3}+k_{4}+k_{-1}+k_{-2}+k_{-3}+k_{-4}+k_{p}+k_{x}+k_{exo}
\end{equation}
\begin{eqnarray}
\gamma & = & k_{1}k_{2}+k_{1}k_{3}+k_{2}k_{3}+k_{1}k_{4}+k_{2}k_{4}+k_{3}k_{4}+k_{exo}(k_{1}+k_{2}+k_{3}+k_{4} \nonumber \\ & + & k_{x} + k_{-1}+k_{-2}+k_{-3}+k_{-4})+ k_{p}(k_{1}+k_{2}+k_{3}+k_{4}+k_{-1}+k_{-2} \nonumber \\ &+& k_{-3}+k_{-4})+k_{x}(k_{2}+k_{3}+k_{4}+k_{-1}+k_{-2}+k_{-3})+ k_{-1}k_{-2} \nonumber \\ &+& 
k_{-1}k_{-3}+ k_{-2}k_{-3}+k_{-1}k_{-4}+k_{-2}k_{-4}+k_{-3}k_{-4} + k_{1}(k_{-2}+k_{-3})\nonumber \\ &+& 
k_{2}(k_{-3}+k_{-4})+k_{3}(k_{-1}+k_{-4})+k_{4}(k_{-1}+k_{-2}+k_{-4})
\end{eqnarray}
\begin{eqnarray}
\delta &=& k_{1}k_{2}k_{3}+k_{1}k_{2}k_{4}+k_{1}k_{3}k_{4}+k_{2}k_{3}k_{4}+k_{-1}k_{-2}k_{-3}+k_{-1}k_{-2}k_{-4}\nonumber \\ &+& k_{-1}k_{-3}k_{-4}+k_{-2}k_{-3}k_{-4}+k_{exo}(k_{1}k_{2}+k_{1}k_{3}+k_{2}k_{3}+k_{1}k_{4}+k_{2}k_{4}\nonumber \\ &+& k_{3}k_{4}+k_{-1}k_{-2}+k_{-1}k_{-3}+k_{-2}k_{-3}+k_{-1}k_{-4}+k_{-2}k_{-4}+k_{-3}k_{-4}) \nonumber \\
&+& k_{exo}k_{x}(k_{2}+k_{3}+k_{4}+k_{-1}+k_{-2}+k_{-3})+k_{exo}(k_{3}k_{-1}+k_{4}k_{-1}+k_{1}k_{-2} \nonumber \\ &+& k_{4}k_{-2}+k_{1}k_{-3}+k_{2}k_{-3}+k_{2}k_{-4}+k_{3}k_{-4}+k_{4}k_{-4})+k_{p}(k_{1}k_{2}+k_{1}k_{3} \nonumber \\ &+& k_{2}k_{3}+k_{1}k_{4}+k_{2}k_{4}+k_{3}k_{4}+k_{-1}k_{-2}+k_{-1}k_{-3}+k_{-2}k_{-3}+k_{-1}k_{-4}
\nonumber \\ &+& k_{-2}k_{-4}+k_{-3}k_{-4}+k_{3}k_{-1}+k_{4}k_{-1}+k_{1}k_{-2}+k_{4}k_{-2}+k_{1}k_{-3}+k_{2}k_{-3}
\nonumber \\ &+ & k_{2}k_{-4}+k_{3}k_{-4}+k_{4}k_{-4})+k_{x}(k_{2}k_{3}+k_{2}k_{4}+k_{3}k_{4}+k_{-1}k_{-3}+k_{-2}k_{-3}
\nonumber \\ &+& k_{-1}k_{-2}+k_{3}k_{-1}+k_{4}k_{-1}+k_{4}k_{-2}+k_{2}k_{-3})+k_{3}k_{4}k_{-1}+k_{1}k_{4}k_{-2}
\nonumber \\ &+& k_{4}k_{-1}k_{-2}+k_{1}k_{2}k_{-3}+k_{1}k_{-2}k_{-3}+k_{2}k_{3}k_{-4}+k_{2}k_{4}k_{-4}
+k_{3}k_{4}k_{-4} \nonumber \\  &+& k_{3}k_{-1}k_{-4}+k_{4}k_{-1}k_{-4}+k_{4}k_{-2}k_{-4}+k_{2}k_{-3}k_{-4}
\end{eqnarray}
\begin{eqnarray}
\epsilon &=& k_{1}k_{2}k_{3}k_{4} + k_{-1}k_{-2}k_{-3}k_{-4} + k_{exo}(k_{1}k_{2}k_{3}+k_{1}k_{2}k_{4}+k_{1}k_{3}k_{4}+k_{2}k_{3}k_{4} \nonumber \\ &+& k_{-1}k_{-2}k_{-3}+k_{-1}k_{-2}k_{-4}+k_{-1}k_{-3}k_{-4}+k_{-2}k_{-3}k_{-4})+k_{exo}k_{x}(k_{2}k_{3}
\nonumber \\ &+& k_{2}k_{4}+k_{3}k_{4}+ k_{3}k_{-1}+k_{4}k_{-1}+k_{4}k_{-2}+k_{-1}k_{-2}+k_{2}k_{-3}+ \nonumber \\ &+& k_{-1}k_{-3}+ k_{-2}k_{-3})+k_{exo}(k_{3}k_{4}k_{-1}+k_{1}k_{4}k_{-2}+k_{4}k_{-1}k_{-2}+k_{1}k_{2}k_{-3}
 \nonumber \\ & + & k_{1}k_{-2}k_{-3}+k_{2}k_{3}k_{-4}
+ k_{2}k_{4}k_{-4}+k_{3}k_{4}k_{-4}+k_{3}k_{-1}k_{-4}+k_{4}k_{-1}k_{-4} \nonumber \\ &+& k_{4}k_{-2}k_{-4}+k_{2}k_{-3}k_{-4})+k_{p}(k_{1}k_{2}k_{3}+k_{1}k_{2}k_{4}+k_{1}k_{3}k_{4}
+k_{2}k_{3}k_{4}\nonumber \\ &+& k_{-1}k_{-2}k_{-3}+k_{-1}k_{-2}k_{-4}+k_{-1}k_{-3}k_{-4}
+k_{-2}k_{-3}k_{-4}+k_{3}k_{4}k_{-1} \nonumber \\ &+& k_{1}k_{4}k_{-2}+ k_{4}k_{-1}k_{-2}+k_{1}k_{2}k_{-3}+k_{1}k_{-2}k_{-3}+k_{2}k_{3}k_{-4}+k_{2}k_{4}k_{-4}\nonumber \\ &+& k_{3}k_{4}k_{-4}
+ k_{3}k_{-1}k_{-4}+k_{4}k_{-2}k_{-4}+k_{2}k_{-3}k_{-4}+k_{4}k_{-1}k_{-4})+k_{x}(k_{2}k_{3}k_{4}\nonumber \\ &+& k_{-1}k_{-2}k_{-3}+ k_{3}k_{4}k_{-1}+k_{4}k_{-1}k_{-2})+k_{4}k_{-1}k_{-2}k_{-3}+k_{3}k_{4}k_{-1}k_{-4}\nonumber \\ &+& k_{2}k_{3}k_{4}k_{-4}
\end{eqnarray}
\begin{eqnarray}
\zeta & = & k_{exo}(k_{1}k_{2}k_{3}k_{4}+k_{2}k_{3}k_{4}k_{x}+k_{3}k_{4}k_{x}k_{-1}+k_{4}k_{x}k_{-1}k_{-2}+
k_{x}k_{-1}k_{-2}k_{-3}\nonumber \\ &+& k_{2}k_{3}k_{4}k_{-4}+k_{3}k_{4}k_{-1}k_{-4}+k_{4}k_{-1}k_{-2}k_{-4}+
k_{-1}k_{-2}k_{-3}k_{-4}) \nonumber \\ &+& k_{p}(k_{1}k_{2}k_{3}k_{4} +k_{2}k_{3}k_{4}k_{-4}+ k_{3}k_{4}k_{-1}k_{-4}+k_{4}k_{-1}k_{-2}k_{-4} \nonumber\\ &+& k_{-1}k_{-2}k_{-3}k_{-4})
\end{eqnarray}

\noindent{\bf Appendix D}

\begin{equation}
a_{0}=k_{1}k_{2}k_{3}(k_{exo}+k_{p})
\end{equation}
\begin{equation}
a_{1}=k_{1}k_{2}k_{3}
\end{equation}

\begin{eqnarray}
b_{0}=(k_{exo}+k_{p})(k_{2}k_{3}k_{4}+k_{3}k_{4}k_{-1}+k_{4}k_{-1}k_{-2}+k_{-1}k_{-2}k_{-3})
\end{eqnarray}
\begin{eqnarray}
b_{1} &=& k_{2}k_{3}k_{4}+k_{3}k_{4}k_{-1}+k_{4}k_{-1}k_{-2}+k_{-1}k_{-2}k_{-3}+k_{exo}(k_{2}k_{3}+k_{2}k_{4} \nonumber \\ &+& k_{3}k_{4} + k_{-1}k_{-2} +k_{-2}k_{-3}+k_{-3}k_{-1}+k_{3}k_{-1}+k_{4}k_{-1}+k_{2}k_{-3}+k_{4}k_{-2})\nonumber \\ &+& k_{p}(k_{2}k_{3}
+ k_{2}k_{4} +k_{3}k_{4}+k_{-1}k_{-2}+k_{-2}k_{-3}+k_{-3}k_{-1}+k_{3}k_{-1}+k_{4}k_{-1}\nonumber \\ &+& k_{4}k_{-2} +k_{2}k_{-3})
\end{eqnarray}
\begin{eqnarray}
b_{2}&=&k_{2}k_{3}+k_{2}k_{4}+k_{3}k_{4}+(k_{exo}+k_{p})(k_{2}+k_{3}+k_{4}+k_{-1}+k_{-2}+k_{-3}) \nonumber \\ &+& k_{3}k_{-1}+k_{4}k_{-1}+k_{4}k_{-2}+k_{-1}k_{-2}+k_{2}k_{-3}+k_{-1}k_{-3}+k_{-2}k_{-3}
\end{eqnarray}
\begin{equation}
b_{3}=k_{2}+k_{3}+k_{4}+k_{exo}+k_{p}+k_{-1}+k_{-2}+k_{-3}
\end{equation}
\begin{equation}
b_{4}=1
\end{equation}

\begin{equation}
c_{0}=k_{x}(k_{2}k_{3}k_{4}+k_{-1}k_{-2}k_{-3}+k_{3}k_{4}k_{-1}+k_{4}k_{-1}k_{-2})
\end{equation}
\begin{eqnarray}
c_{1}&=&k_{x}(k_{2}k_{3}+k_{2}k_{4}+k_{3}k_{4}+k_{3}k_{-1}+k_{4}k_{-1}+k_{4}k_{-2}+k_{-1}k_{-2}+k_{2}k_{-3}\nonumber \\&+&k_{-1}k_{-3}
+k_{-2}k_{-3})
\end{eqnarray}
\begin{equation}
c_{2}=k_{x}(k_{2}+k_{3}+k_{4}+k_{-1}+k_{-2}+k_{-3})
\end{equation}
\begin{equation}
c_{3}=k_{x}
\end{equation}

\begin{eqnarray}
d_{0} &=& k_{exo}(k_{1}k_{2}k_{3}+k_{2}k_{3}k_{x}+k_{3}k_{x}k_{-1}+k_{x}k_{-1}k_{-2}+k_{2}k_{3}k_{-4}
+k_{3}k_{-1}k_{-4} \nonumber \\ &+& k_{-1}k_{-2}k_{-4})+k_{p}(k_{1}k_{2}k_{3}+k_{2}k_{3}k_{-4}+k_{3}k_{-1}k_{-4}
+k_{-1}k_{-2}k_{-4})
\end{eqnarray}
\begin{eqnarray}
d_{1}&=& k_{1}k_{2}k_{3}+k_{-1}k_{-2}k_{-4}+k_{exo}(k_{1}k_{2}+k_{1}k_{3}+k_{2}k_{3}+k_{3}k_{-1} + k_{1}k_{-2}\nonumber \\ &+&k_{2}k_{-4} + k_{3}k_{-4}+k_{-1}k_{-2}+k_{-1}k_{-4}+k_{-2}k_{-4})+k_{exo}k_{x}(k_{2}+k_{3} \nonumber \\ &+& k_{-1}+k_{-2})
+k_{p}(k_{1}k_{2}+k_{1}k_{3}+k_{2}k_{3}+k_{3}k_{-1}+k_{1}k_{-2}+k_{2}k_{-4} \nonumber \\ &+& k_{3}k_{-4}+k_{-1}k_{-2}+k_{-1}k_{-4}
+k_{-2}k_{-4})+k_{x}(k_{2}k_{3}+k_{3}k_{-1}+k_{-1}k_{-2})\nonumber\\ &+& k_{2}k_{3}k_{-4}+k_{3}k_{-1}k_{-4}
\end{eqnarray}
\begin{eqnarray}
d_{2}&=&k_{1}k_{2}+k_{1}k_{3}+k_{2}k_{3}+k_{exo}(k_{1}+k_{2}+k_{3}+k_{x}+k_{-1}+k_{-2}+k_{-4}) \nonumber \\ &+&k_{p}(k_{1}+k_{2}+k_{3}+k_{-1}+k_{-2}+k_{-4})+k_{x}(k_{2}+k_{3}+k_{-1}+k_{-2})\nonumber \\&+&k_{3}k_{-1}+k_{1}k_{-2}+k_{-1}k_{-2}+ k_{2}k_{-4}+k_{-1}k_{-4}+k_{-2}k_{-4}+k_{3}k_{-4}
\end{eqnarray}
\begin{equation}
d_{3}=k_{1}+k_{2}+k_{3}+k_{exo}+k_{p}+k_{x}+k_{-1}+k_{-2}+k_{-4}
\end{equation}
\begin{equation}
d_{4}=1
\end{equation}

\begin{equation}
e_{0}=k_{-1}k_{-2}k_{-3}(k_{exo}+k_{p})
\end{equation}
\begin{equation}
e_{1}=k_{-1}k_{-2}k_{-3}
\end{equation}

\begin{eqnarray}
f_{0}&=&k_{1}k_{2}k_{3}k_{4}+k_{-1}k_{-2}k_{-3}k_{-4}+k_{x}(k_{2}k_{3}k_{4}+k_{3}k_{4}k_{-1}+k_{4}k_{-1}k_{-2}\nonumber \\ &+&k_{-1}k_{-2}k_{-3}) + k_{3}k_{4}k_{-1}k_{-4}
+k_{4}k_{-1}k_{-2}k_{-4}+k_{2}k_{3}k_{4}k_{-4}
\end{eqnarray}
\begin{eqnarray}
f_{1}&=&k_{1}k_{2}k_{3}+k_{1}k_{2}k_{4}+k_{1}k_{3}k_{4}+k_{2}k_{3}k_{4}+k_{3}k_{4}k_{-1}+k_{1}k_{4}k_{-2}+
k_{4}k_{-1}k_{-2}\nonumber \\ &+&k_{1}k_{2}k_{-3}+k_{1}k_{-2}k_{-3}+k_{-1}k_{-2}k_{-3}+k_{2}k_{3}k_{-4}+k_{2}k_{4}k_{-4}+k_{3}k_{4}k_{-4}
\nonumber \\ &+&
k_{3}k_{-1}k_{-4} + k_{4}k_{-1}k_{-4}+k_{4}k_{-2}k_{-4}+k_{-1}k_{-2}k_{-4}+k_{2}k_{-3}k_{-4} \nonumber \\ &+&
k_{-1}k_{-3}k_{-4}+k_{-2}k_{-3}k_{-4}  +k_{x}(k_{2}k_{3}+k_{2}k_{4}+k_{3}k_{4}+k_{3}k_{-1}+k_{4}k_{-1} 
\nonumber \\&+& k_{4}k_{-2} + k_{-1}k_{-2}+k_{2}k_{-3}+k_{-1}k_{-3}+k_{-2}k_{-3})
\end{eqnarray}
\begin{eqnarray}
f_{2} &=& k_{1}k_{2}+k_{1}k_{3}+k_{2}k_{3}+k_{1}k_{4}+k_{2}k_{4}+k_{3}k_{4}+k_{3}k_{-1}+k_{4}k_{-1}+k_{1}k_{-2}
\nonumber \\ &+&k_{4}k_{-2} + k_{-1}k_{-2}+k_{1}k_{-3}+k_{2}k_{-3}+k_{-1}k_{-3}+k_{-2}k_{-3}+k_{2}k_{-4} \nonumber \\ &+& k_{3}k_{-4}+k_{4}k_{-4}+ k_{-1}k_{-4}+k_{-2}k_{-4}+k_{-3}k_{-4}+k_{x}(k_{2}+k_{3}+k_{4} \nonumber \\ &+& k_{-1}+k_{-2}+k_{-3}) 
\end{eqnarray}
\begin{equation}
f_{3}=k_{1}+k_{2}+k_{3}+k_{4}+k_{x}+k_{-1}+k_{-2}+k_{-3}+k_{-4}
\end{equation}
\begin{equation}
f_{4}=1
\end{equation}

\begin{equation}
g_{0}=k_{p}(k_{2}k_{3}k_{4}+k_{3}k_{4}k_{-1}+k_{4}k_{-1}k_{-2}+k_{-1}k_{-2}k_{-3})
\end{equation}
\begin{eqnarray}
g_{1}&=&k_{p}(k_{2}k_{3}+k_{2}k_{4}+k_{3}k_{4}+k_{3}k_{-1}+k_{4}k_{-1}+k_{4}k_{-2}+k_{-1}k_{-2}+k_{2}k_{-3}\nonumber \\&+&k_{-1}k_{-3}
+k_{-2}k_{-3})
\end{eqnarray}
\begin{equation}
g_{2}=k_{p}(k_{2}+k_{3}+k_{4}+k_{-1}+k_{-2}+k_{-3})
\end{equation}
\begin{equation}
g_{3}=k_{p}
\end{equation}


\end{document}